\documentclass[preprint,showpacs,preprintnumbers,amsmath,amssymb]{revtex4}
\usepackage{graphicx,epsfig,dcolumn,bm,epic,eepic,float}%
\usepackage{amsmath}
\usepackage{latexsym}
\usepackage{color}
\usepackage{cancel}
\usepackage{makeidx,shortvrb,latexsym}
\usepackage{slashed}
\begin{document}
\unitlength 1 cm
\newcommand{\nn}{\nonumber}
\newcommand{\vk}{\vec k}
\newcommand{\vp}{\vec p}
\newcommand{\vq}{\vec q}
\newcommand{\vkp}{\vec {k'}}
\newcommand{\vpp}{\vec {p'}}
\newcommand{\vqp}{\vec {q'}}
\newcommand{\bk}{{\bf k}}
\newcommand{\bp}{{\bf p}}
\newcommand{\bq}{{\bf q}}
\newcommand{\br}{{\bf r}}
\newcommand{\bR}{{\bf R}}
\newcommand{\up}{\uparrow}
\newcommand{\down}{\downarrow}
\newcommand{\fns}{\footnotesize}
\newcommand{\ns}{\normalsize}
\newcommand{\cdag}{c^{\dagger}}
\title {The analysis of Drell-Yan lepton pair production in the P-P(\={P}) colliders using different
angular ordering constraints and $k_t$-factorization approach}
\author{$M. \; Modarres$ }
\altaffiliation {Corresponding author, Email:
mmodares@ut.ac.ir, Tel:+98-21-61118645, Fax:+98-21-88004781.}
\author{$R. \; Taghavi$ }
\affiliation {Department of Physics, University of $Tehran$,
1439955961, $Tehran$, Iran.}
\author{$R. \; Aminzadeh \;Nik$ }
\author{$R.\; Kord \; Valeshabadi$ }
\affiliation {Department of Physics, University of $Tehran$,
1439955961, $Tehran$, Iran.} \ \
\begin{abstract}
In this work, the P-P(\={P}) Drell-Yan lepton pair production (DY) differential cross sections at hadrons
colliders, such as LHC and $\;$ TEVATRON, are studied in the ${k_{t}}$-factorization framework. In
order to take into account the transverse momenta of   incoming
partons, we use the unintegrated parton distribution functions
(UPDF) of  Kimber et al  (KMR) and  Martin et al (MRW) in the
leading order (LO) and  next-to-leading-order (NLO) levels  with the
input MMHT2014 PDF libraries. Based on the different off-shell
partonic matrix elements, we analyze the behaviors of DY differential
cross sections with respect to the  invariant mass, the transverse
momentum and the rapidity as well as the specific angular correlation
between the produced leptons. The numerical results are compared with the
experimental  data, in different energies, which are reported by
various collaborations, such as CDF, CMS, ATLAS and LHCb. It is
shown that the NLO-MRW and KMR schemes predict closer results to the
data  compared to the LO-MRW, since we do not have fragmentation. It
is demonstrated  that while the $ q^* + \bar q^* \to {{{\gamma ^*}}
\mathord{\left/
 {\vphantom {{{\gamma ^*}} {Z + g \to {l^ + } + {l^ - }}}} \right.
 \kern-\nulldelimiterspace} {Z + g \to {l^ + } + {l^ - }}} + g$
 sub-process has a negligible contribution, it has a sizable effect 
in the low dilepton transverse momentum. In agreement with the NNLO pQCD report (PYTHIA, SHERPA, etc),   by including the higher order perturbative
contributions   the better results are archived. On the other hand as
the scale of energy increases, for the LHC energies, the Compton
  sub-process, i.e.,  $q^* + {g^*} \to {{{\gamma ^*}} \mathord{\left/
 {\vphantom {{{\gamma ^*}} {Z \to {l^ + } + {l^ - } + q}}} \right.
 \kern-\nulldelimiterspace} {Z \to {l^ + } + {l^ - } + q}}$, has the largest contribution to the differential cross section in the most
  intervals of some observables, as it is expected. 
 The variation of the differential cross-section with respect to
the various variables such as the invariant mass, the center of mass energy, etc are discussed. 
In order to validate our results, we   also consider  the strong ordering constraint and the KaTie parton-level event generator.
\end{abstract}

\pacs{12.38.Bx, 13.85.Qk, 13.60.-r
\\ \textbf{Keywords:}  unintegrated parton distribution function,
${k_{t}}$-facorization, Drell-Yan lepton pair production.
\\
\\ (In the online journal, the figures are colored)} \maketitle

\section{Introduction}
Traditionally,  the Dokshitzer-Gribov-Lipatov-Altarelli-Parisi
(DGLAP) evolution equations \cite{DGLAP1,DGLAP2,DGLAP3,DGLAP4}
approach is used to obtain the quark, anti-quark and gluon
densities, i.e., the parton distribution functions (PDF),
$a(x,\mu^2)$. These functions   depend on the Bjorken variable x
and the hard scale ${\mu ^2}$, and can be easily used in   the
collinear QCD factorization formalisms. In these DGLAP evolution approaches,    the
transverse momentum ($k_t$)  components of partons are integrated
over and there is not any degree of freedom  for the initial gluon 
radiations and the transverse momentum ($k_t$) of the partons in the
PDF.  But, the outcome of hadron-hadron colliders at high energies
indicates that the explicit inclusion of  the intrinsic transverse
momentum of   initial hadron constituents is important to get  
accurate results and predictions. Therefore, the important inputs are the
transverse momentum dependent parton distribution functions   or the
so-called "unintegrated" PDF (UPDF).

Theoretically, various methods are proposed to generate these
fundamental quantities, i.e., UPDF, and among them the BFKL
\cite{BFKL1,BFKL2,BFKL3,BFKL4,BFKL5} (which is valid for the small x
and the scale $k_t^2$) and  CCFM
\cite{CCFM1,CCFM2,CCFM3,CCFM4,CCFM5} (which is applicable at  both
the small and large x, the scales $k_t^2$ and $\mu^2$) evolution
equations are considered extensively. Nevertheless,  the CCFM
approach is both mathematically and numerically more complicated and
time consuming.

Recently,   Kimber et al \cite{KMR} and Martin et al \cite{MRW}
proposed the KMR and MRW formalisms, respectively,  in the leading
order
 (LO) and next-to-leading order (NLO) levels. These formalisms were  extensively used to extract
 the UPDF, $f(x,k^2_t,\mu^2)$, from the ordinary PDF based on the ${k_{t}}$-factorization
 approach of pQCD as well as probing the partonic structures of hadrons 
\cite{10,11,12,13,14}. The main difference between these two
approaches is turned back into the various types of  imposing  the
angular ordering constraints (AOC). These formalisms were analyzed
by us  to calculate the proton structure functions and the different
hadron-hadron differential cross sections in the  references
\cite{15,16,17,18,19}.

The analysis of Drell-Yan lepton pair production (DY) in the
hadron-hadron collisions
 at high energies is the subject of intense studies \cite{20,21,22,23,24,25,26,27,phi3}, since it provides an ideal
 ground for testing the QCD predictions \cite{book1,book2,Klasen,28}. Many experimental groups like the CDF,
 CMS, ATLAS, D0 and LHCb collaborations \cite{20,21,22,23,24,25,26,phi3} in the  available energies of the Tevatron and   LHC colliders,  try to
 compare the experimental measurements of these DY events to the corresponding theoretical predictions from
 the pQCD and the parton level Monte Carlo programs, such as
  ResBos \cite{ResBos1}, DYNNLO \cite{DYNNLO1} and  POWHEG+PYTHIA \cite{pythia1} event generators.
The  ResBos method simulates the vector-boson
production and its decay,
    using a resumed treatment of the soft-gluon emissions at the NLO-logarithm
  (NNLL) and the $\gamma ^*$
 and Z/$\gamma ^*$ contributions are simulated at the NLO accuracies.
 The DYNNLO approach  is a parton level Monte Carlo program that computes the
 cross sections for the vector boson production in the P-P(\={P}) collisions up to the NNLO in the pQCD theory.
 The PYTHIA program  generates the LO QCD interactions via its parton shower algorithms.
In the recent investigation \cite{26} the distribution of dilepton
 transverse momentum and angular variable $\phi _\eta ^*$
 were calculated perturbatively at $\sqrt s  = 8$ TeV using the ResBos Monte Carlo
 generator  at the  NNLO accuracy
  and compared to the ATLAS data. Although, the results at low values of
  $p_T$ and $\phi _\eta ^*$   show a good agreement
   with data, but  this is not the case  at high values of $p_T$ and $\phi _\eta ^*$.

In the present report, it is intended to calculate the DY differential cross sections
 based on the KMR and MRW ${k_{t}}$-factorization approaches
 using the corresponding off-shell transition amplitudes. We consider the three sub-processes
 namely, (1)
$q^* + \bar q^* \to {{{\gamma ^*}} \mathord{\left/
 {\vphantom {{{\gamma ^*}} {Z \to {l^ + } + {l^ - }}}} \right.
 \kern-\nulldelimiterspace} {Z \to {l^ + } + {l^ - }}}$ and (2) $q^* + {g^*} \to {{{\gamma ^*}} \mathord{\left/
 {\vphantom {{{\gamma ^*}} {Z \to {l^ + } + {l^ - } + q}}} \right.
 \kern-\nulldelimiterspace} {Z \to {l^ + } + {l^ - } + q}}$ and (3) $ q^* + \bar q^* \to {{{\gamma ^*}} \mathord{\left/
 {\vphantom {{{\gamma ^*}} {Z + g \to {l^ + } + {l^ - }}}} \right.
 \kern-\nulldelimiterspace} {Z + g \to {l^ + } + {l^ - }}} + g$
  at the LO and NLO levels, respectively.
  The dependence of the DY differential cross sections on  the dilepton transverse momentum, the invariant mass
 and the rapidity distributions as well as the angular correlation between the produced
 leptons are calculated in the above frameworks and compared with the experimental data developed  by the CDF,
 CMS, ATLAS and LHCb collaborations in both the Tevatron and   LHC
 energies. The Harland-Lang et al. (MMHT2014) PDF libraries \cite{mmht2014} in the LO
 for both the  KMR \cite{KMR} and MRW \cite{MRW} formalisms and the NLO for NLO-MRW   approaches are considered. These calculations
 are performed for the off-shell   incoming partons. 
 
 We should also
 point out here that, recently, the consideration of various
 angular ordering, as well as the generation of different UPDF becomes
 the subject of several reports \cite{re1,re2}. In the reference \cite{re1}, it is pointed out that        the KMR UPDF,  which are generated using   the differential  and    integral approaches, give different results in the region where $k_t > \mu$. Therefore it is  concluded  that, the integral form of the KMR UPDF   gives the correct result, while  for the application of    differential form, one should use the cutoff dependent PDF. On the other hand, in the reference \cite{Guiot:2019vsm},  the above idea  is rejected, and   a new term is added to the Sudakov form factor via a Heaviside step function, to set the Sudakov form factor equal to $1$ in the region where $k_t > \mu$.  Further, it is claim that the above two   forms of   KMR approach give the same result, i.e., there is no need to introduce cutoff dependent PDF. Finally, in the reference \cite{Guiot:2019vsm}, by referring to this report, i.e.,   \cite{Guiot:2018kfy} ( in which the predictions of the KMR approach with the  AOC overestimates the data in case of the heavy quark production), it is   suggested that the above problem is due to the freedom of parton to have transverse momentum larger than $\mu$, and concluded that it is much suitable to use the KMR UPDF with the strong ordering constraint (SOC), which harshly cuts the transverse momentum in  $k_t > \mu$ region. Because of the above statements, and   all of the problems appear in   $k_t > \mu$, we   compute the DY differential cross sections with respect to $M^{ll}$ and $p_T^{ll}$ using the SOC   KMR UPDF to check the sensitivity of our results in this region (see the figures 1 (panel f) and 2 (panel d)). One should also note that, as it is  discussed
in the reference \cite{Watt:2003mx}, the result of $k_t$-factorization should not be as good as collinear factorization  approaches, in  covering the experimental data, on the other hand, it is more simplistic, considering computer time consuming. 
 
It should be also noted that the Sudakov form factor of the KMR approach does not obey the multiplication law according to the reference \cite{Hautmann:2019biw}, but despite of this fact, it is interesting to point out  that the normalization condition (see the equation (\ref{NOR})) is approximately satisfied in the KMR formalism, which is a critical issue in constructing any new UPDF. 

The $k_t$-factorization  calculations were also performed  by
considering  one or two of the above three sub-processes with the
MSTW2008 PDF
 \cite{lipatov2011,reggeiz1,reggeiz2}. In these works, although the authors declare that they use the KMR
 formalism, but they do not take into account  the factor $1/{k_t^2}$ in the cross section nor in the normalization
 formulas:
  \begin{equation}
xa(x,\mu^2) \simeq \int^{\mu^2} {{dk_t^2}\over{k_t^2}}
f(x,k^2_t,\mu^2).\label{NOR}
\end{equation}
Beside these,  they use different angular ordering conditions with
respect to  the KMR prescriptions (we refer to them as semi-KMR).
However, their results are surprisingly close to the experimental
data. In the above references \cite{reggeiz1,reggeiz2}, it is
claimed that the second and the third of the above
 sub-processes can be omitted by effectively using only the
reggeized  (off-shell) quark approach         in  the first
sub-process. A brief discussion about the result of these reports
   and the comparison with our predictions  are presented in the section III. In the reference  \cite{reggeiz1},
  although the off-shell initial quarks are  used, it is shown that utilizing
  the reggeized model and the effective vertexes guaranteed the gauge invariance of the transition matrix elements (TME). However,
  in our previous work \cite{diphoton}, we showed that using the off-shell initial quarks in the
  $k_t$-factorization dynamics and in the small x regions leads to the gauge invariance of the TME, too.
  
To check the validity of our calculated cross sections the KaTie parton-level event generator \cite{KaTie} is used, in which    the off-shell partonic cross
sections  are taken care of, and gives the hadronic cross section with desirable accuracy. However,
we are not intended to solely show the result of cross section as it can be simply done with the KaTie parton-level event generator \cite{KaTie} (see the figures 1 (panel f) and 2 (panel h)). Indeed, our intention from one side is to check the effects of different impositions of the cutoff $\Delta$, which is additionally imposed on the quark radiation term in the KMR approach.  On the other side, we also want to check the other forms of the DGLAP based UPDF, i.e. NLO-MRW, in which Martin et. al. used the virtuality $k^2 = k_t^2/(1-z)$. The UPDF  of this form are rarely investigated in the phenomenological applications of the $k_t\textrm{-factorization}$. We  include the sub-process $q^*+\bar{q}^*\rightarrow \gamma^*/Z\rightarrow l^++l^-+g$ 
which usually is neglected,   e.g. \cite{lipatov2011}. In the others works, including those that are cited in our report  \cite{lipatov2011}, incorrectly, the combinations of KMR and MRW formalisms are used and they forget about the importance of the normalization constraint (the equation (1)) on the UPDF. This point is discussed in details in the reference \cite{diphoton}. One should note that the KMR prescription is a semi-NLO. Another important item is the fragmentation effect, which does not present in the processes that are discussed in our report. Because of that, as it is explained in the paper, the KMR and NLO-MRW procedures demonstrate better agreement to the experimental data.

 The outline of our paper is as following. In
the section II, we briefly present  the basic cross section formulas
of ${k_{t}}$-factorization (\ref{CS}) approach and the derivation of
input UPDF (\ref{U}). In the section III we present numerical
calculations (\ref{1}), results presentations (\ref{2}) and
discussions (\ref{2}). The section IV expresses our conclusions.
\section{The Theoretical framework of DY}
\subsection{The $k_t$-factorization  cross section formulas}
\label{CS}
 Our DY differential cross section calculations  are based
on the $k_t$-factorization in   the KMR and MRW UPDF \cite{KMR,MRW}
approaches. Therefore in this section, we describe the theoretical framework
of these approaches as well as the corresponding matrix elements
(also see the appendix \ref{a}). As we pointed out in the
introduction, we include all the sub-processes contributions up to
the $\alpha\alpha_s$ levels, namely: $q^* + \bar q^* \to {{{\gamma ^*}}
\mathord{\left/
 {\vphantom {{{\gamma ^*}} {Z \to {l^ +  } + {l^ - }}}} \right.
 \kern-\nulldelimiterspace} {Z \to {l^ + } + {l^ - }}}$,
$q^* + {g^*} \to {{{\gamma ^*}} \mathord{\left/
 {\vphantom {{{\gamma ^*}} {Z \to {l^ + } + {l^ - } + q}}} \right.
 \kern-\nulldelimiterspace} {Z \to {l^ + } + {l^ - } + q}}$ and
$q^* + \bar q^* \to {{{\gamma ^*}} \mathord{\left/
 {\vphantom {{{\gamma ^*}} {Z + g \to {l^ + } + {l^ - }}}} \right.
 \kern-\nulldelimiterspace} {Z + g \to {l^ + } + {l^ - }}} + g$.
From the kinematical point of view, if we show the four-momenta of
the incoming protons (partons) by ${P^{(1)}}$ (${k^{(1)}}$) and
${P^{(2)}}$ (${k^{(2)}}$) and neglect their masses, then in the
proton center of mass framework we have:
 \begin{equation}
{P^{(1)}} = \frac{{\sqrt s }}{2}(1,0,0,1), \ \ \  {P^{(2)}} =
\frac{{\sqrt s }}{2}(1,0,0, - 1),
 \end{equation}
where ${\sqrt s }$ is the total center of mass energy. In the high
energy and the leading-log-approximation kinematics, the
corresponding partons four-momenta can be written in terms of their
transverse momenta ${k_{1t}}$ and ${k_{2t}}$ and  the fraction
($x_i$) of the incoming protons momentum as:
 \begin{equation}
{k_1} = {x_1}{P^{(1)}} + {k_{1t}}, \ \ \  {k_2} = {x_2}{P^{(2)}} +
{k_{2t}}.
 \end{equation}
There are some relations for the above three  sub-processes   due to
the energy-momentum conservation law as following:
 \begin{equation}
{k_{1t}} + {k_{2t}} = {p_{1t}} + {p_{2t}},
 \end{equation}
 \begin{equation}
{x_1} = \frac{1}{{\sqrt s }}({m_{1t}}{e^{{y_1}}} +
{m_{2t}}{e^{{y_2}}}),
 \end{equation}
 \begin{equation}
{x_2} = \frac{1}{{\sqrt s }}({m_{1t}}{e^{ - {y_1}}} + {m_{2t}}{e^{ -
{y_2}}}).
 \end{equation}
for the first sub-process and
 \begin{equation}
{k_{1t}} + {k_{2t}} = {p_{1t}} + {p_{2t}} + {p_{3t}},
\end{equation}
 \begin{equation}
 {x_1} = \frac{1}{{\sqrt s }}({m_{1t}}{e^{{y_1}}} + {m_{2t}}{e^{{y_2}}} + {m_{3t}}{e^{{y_3}}}),
 \end{equation}
 \begin{equation}
{x_2} = \frac{1}{{\sqrt s }}({m_{1t}}{e^{ - {y_1}}} + {m_{2t}}{e^{ -
{y_2}}} + {m_{3t}}{e^{ - {y_3}}}) \label{9}.
 \end{equation}
for the second and third sub-processes, where   ${p_{it}}$,
${y_i}$ and ${m_{it}}$ are the transverse momenta,  
  the rapidities and the transverse masses ($m_{it}^2
= m_i^2 + p_{it}^2$) of the produced particles, ($i$=1 and 2 for
leptons and $i$=3 for (anti-)quark or gluon), respectively.

To  calculate the matrix elements squared in the $k_t$-factorization
framework \cite{43}, the summation over the incoming off-shell gluon
polarizations is carried out as:
 \begin{equation}
\sum {{\varepsilon ^\mu }{\varepsilon ^\nu } = k_{2t}^\mu }
k_{2t}^\nu /k_{2t}^2,
 \end{equation}
where ${k_{t}}$ is the gluon transverse momentum. For the
off-shell quarks spinors with momentum $k$ (after imposing the Sudakov decomposition in the high
energy    and the leading-log-approximation kinematics \cite{n1}), we have $\sum {u(k)\bar u(k)}  \simeq {x\hat P \over k_t^2}$,
where $x$ represents the  fractional longitudinal momentum of
proton, see the references  \cite{diphoton,n2,n3}. Also, the effective vertices are used to calculate the Feynman amplitudes to test and ensure the gauge  invariance of the different matrix elements \cite{LLIP1,LLIP2p,LLIP2}. It is worth to point out that a similar technique is also developed, using the Slavnov-Taylor identities by the means of the helicity amplitude \cite{Kutak0,Kutak1,Kutak2} and being  checked  against those obtained by usage of Lipatov’s effective action \cite{LLIP1,LLIP2p,LLIP2}, as we pointed out.

To calculate the differential cross sections of  DY, according to
the ${k_{t}}$-factorization theorem, for $2 \to 2$ sub-process we
have:
\[\sigma_1  = \sum\limits_q {\int {\frac{1}{{16\pi {{({x_1}{x_2}s)}^2}}}} } {\left| {\cal M}_1^{\gamma^*}+ {\cal M}_1^{Z} \right|^2} \times\]
 \begin{equation}
 {f_q}({x_1},k_{1t}^2,{\mu ^2}){f_{\bar{q}}}({x_2},k_{2t}^2,{\mu ^2})\frac{{d{k _{1t}^2}}}{{k_{1t}^2}}\frac{{d{k _{2t}^2}}}{{k_{2t}^2}}dp_{1t}^2dp_{2t}^2
d{y_1}d{y_2}\frac{{d{\phi _1}}}{{2\pi }}\frac{{d{\phi _2}}}{{2\pi
}},
 \end{equation}
and for $2 \to 3$ sub-process one finds:
\[\sigma_{2(3)}  = \sum\limits_q {\int {\frac{1}{{256{\pi ^3}{{({x_1}{x_2}s)}^2}}}} } {\left| {\cal M}_{2(3)}^{\gamma^*}+ {\cal M}_{2(3)}^{Z} \right|^2} \times \]
 \begin{equation}
{f_q}({x_1},k_{1t}^2,{\mu ^2}){f_{g(\bar{q})}}({x_2},k_{2t}^2,{\mu
^2})\frac{{d{k _{1t}^2}}}{{k_{1t}^2}}\frac{{d{k
_{2t}^2}}}{{k_{2t}^2}}dp_{1t}^2dp_{2t}^2d{y_1}d{y_2}d{y_3}\frac{{d{\phi
_1}}}{{2\pi }}\frac{{d{\phi _2}}}{{2\pi }}\frac{{d{\psi _1}}}{{2\pi
}}\frac{{d{\psi _2}}}{{2\pi }},
 \end{equation}
where ${{f_q}({x_i},k_{it}^2,{\mu ^2})}$ are the UPDF, which depend
on the two hard scales, ${k_{t}^2}$ and  ${\mu ^2}$, and they can be
written in terms of    the usual PDF. As we pointed out in the
introduction, in the present calculations, the MMHT2014 PDF
\cite{mmht2014} is used for calculating the UPDF. In the above
formula, ${\cal M}_i^j$  are the off-shell matrix elements which are
presented for the three different sub-processes in the appendix A.
Note that when we squared the matrix element of each three
sub-processes we get the interference effect between $\gamma^*$ and
$Z$ production, which will be discussed in the section III.  The
azimuthal angles of the initial partons and the produced leptons are
presented by ${\phi _1}$ and  ${\phi _2}$, and ${\psi _1}$ and ${\psi _2}$, respectively.
Then the total cross section can be written as:
 \begin{equation}
 \sigma_{Total}=\sigma_1+\sigma_2+\sigma_3.
\end{equation}
 To calculate the
UPDF of (anti-)quarks and gluons in a proton, we apply the LO KMR,
LO MRW and NLO-MRW approaches \cite{KMR,MRW}. In the following each
of them will be described.
\subsection{The KMR and MRW UPDF}
\label{U}
 In the KMR method the UPDF of each parton, which means the
probability to find a parton with transverse momentum $k_t$ and
fractional momentum $x$ at hard scale $\mu^2$ are given by:
 \begin{equation}
f_a(x,k_t^2,\mu^2) = T_a(k_t^2,\mu^2)\sum_{b=q,g} \left[
{\alpha_S(k_t^2) \over 2\pi} \int^{1-\Delta}_{x} dz P_{ab}^{(0)}(z)
b\left( {x \over z}, k_t^2 \right) \right] , \label{eq56}
    \end{equation}
where ${T_a(k_t^2,\mu^2)}$ is
  \begin{equation}
T_a(k_t^2,\mu^2) = exp \left( - \int_{k_t^2}^{\mu^2} {\alpha_S(k^2)
\over 2\pi} {dk^{2} \over k^2} \sum_{b=q,g} \int^{1-\Delta}_{0} dz'
P_{ab}^{(0)}(z') \right). \label{eq5}
    \end{equation}
which is the familiar Sudakov survival form factor
 and limits the emissions of partons between
${k_t^{2}}$ and $\mu^2 $  scales \cite{KMR,MRW}. ${P_{ab}^{(0)}}(z)$
are the usual LO splitting functions. In this formula the
angular-ordering constraint (AOC)
\cite{CCFM1,CCFM2,CCFM3,CCFM4,CCFM5,44,45}, $\Delta$, is applied in
the upper limit of the integration, which is an infrared cutoff to
prevent the soft gluon singularities arise from the splitting
functions and defined as:
 \begin{equation}
  \Delta=   {k_t \over \mu + k_t}.
  \end{equation}
Note that this constraint is imposed on both quark and gluon
radiations.  ${b\left( {x \over z}, k_t^2 \right)}$  are the LO PDF,
and in this work they are taken from the MMHT2014 libraries
\cite{mmht2014}.

To determine the UPDF we also apply the MRW prescription which is
similar to the KMR formalism, but  the AOC only acts on the terms
which include the on shell gluon emissions. For the quarks  and the
gluons they take the following forms:
$$
f_q^{LO}(x,k_t^2,\mu^2)= T_q(k_t^2,\mu^2) {\alpha_S(k_t^2) \over
2\pi} \int_x^1 dz \left[ P_{qq}^{(0)}(z) {x \over z} q \left( {x
\over z} , k_t^2 \right) \Theta \left( {\mu \over \mu + k_t}-z
\right) \right.
    $$
    \begin{equation}
\left. + P_{qg}^{(0)}(z) {x \over z} g \left( {x \over z} , k_t^2
\right) \right],
    \end{equation}
with
    \begin{equation}
T_q(k_t^2,\mu^2) = exp \left( - \int_{k_t^2}^{\mu^2} {\alpha_S(k^2)
\over 2\pi} {dk^{2} \over k^2} \int^{z_{max}}_{0} dz'
P_{qq}^{(0)}(z') \right),    \end{equation}
  and
    $$
f_g^{LO}(x,k_t^2,\mu^2)= T_g(k_t^2,\mu^2) {\alpha_S(k_t^2) \over
2\pi} \int_x^1 dz \left[ P_{gq}^{(0)}(z) \sum_q {x \over z} q \left(
{x \over z} , k_t^2 \right)
    \right.$$
    \begin{equation}
\left. + P_{gg}^{(0)}(z) {x \over z} g \left( {x \over z} , k_t^2
\right) \Theta \left( {\mu \over \mu + k_t}-z \right)
    \right],
    \end{equation}
with
    \begin{equation}
T_g(k_t^2,\mu^2) = exp \left( - \int_{k_t^2}^{\mu^2} {\alpha_S(k^2)
\over 2\pi} {dk^{2} \over k^2}
    \left[ \int^{z_{max}}_{z_{min}} dz' z' P_{gg}^{(LO)}(z')
+ n_f \int^1_0 dz' P_{qg}^{(0)}(z') \right] \right),
    \end{equation}
respectively. In the above equations, ${{z_{max}=1-{z_{min}}}= {\mu
\over \mu + k_t}}$ \cite{watt}.

By expanding MRW to the NLO level, we have:
    $$
f_a^{NLO}(x,k_t^2,\mu^2)= \int_x^1 dz T_a \left( k^2={k_t^2 \over
(1-z)}, \mu^2 \right) {\alpha_S(k^2) \over 2\pi}
    \sum_{b=q,g} \tilde{P}_{ab}^{(0+1)}(z)
    $$
    \begin{equation}
\times b^{NLO} \left( {x \over z} , k^2 \right) \Theta \left(
1-z-{k_t^2 \over \mu^2} \right).
    \label{eq11}
    \end{equation}
In this formalism the Sudakov form factor is defined as:
  \begin{equation}
T_q(k^2,\mu^2) = exp \left( - \int_{k^2}^{\mu^2} {\alpha_S(q^2)
\over 2\pi} {dq^{2} \over q^2} \int^1_0 dz' z' \left[
\tilde{P}_{qq}^{(0+1)}(z') + \tilde{P}_{gq}^{(0+1)}(z') \right]
\right),
        \end{equation}
    \begin{equation}
T_g(k^2,\mu^2) = exp \left( - \int_{k^2}^{\mu^2} {\alpha_S(q^2)
\over 2\pi} {dq^{2} \over q^2} \int^1_0 dz' z' \left[
\tilde{P}_{gg}^{(0+1)}(z') + 2n_f\tilde{P}_{qg}^{(0+1)}(z') \right]
\right) .
\end{equation}
 The higher order  splitting functions
are presented in the appendix B.
\section{Numerical results and discussions}
\subsection{Numerical calculations}
\label{1} In this section, we present the kinematics and theoretical
inputs of our calculations. First, we calculate the UPDF based on the
different  ${k_{t}}$-factorization schemes by using two different
methods, i.e., KMR and MRW. Through our  calculations, we set the
renormalization and factorization scales to be equal to $\mu _R =
\mu _F = \zeta M$, in which,
$M=\sqrt{2p_{1t}p_{2t}[\cosh(y_1-y_2)-\cos(\phi_1-\phi_2)]}$ is the
invariant mass of produced dilepton and as usual we consider the
default value $\zeta = 1$ \cite{diphoton}. We let this parameter to
vary  from 1/2 to 2, to estimate the scale uncertainties of our
calculations. We also set ${m_Z} = 91.187$ $GeV$ and ${\Lambda
_{QCD}} = 200 MeV$ with ${n_f} = 4$ active quark flavors. Using the
LO coupling constant, we get ${\alpha _s}(M_Z^2) = 0.123$ ($g_W= 0.66$). Second,
with the massless quarks approximation, the calculation of
transition matrix elements squared is carried out, using the small x
approximation presented in the appendix \ref{a}, by the means of
FeynCalc \cite{feyncalc}, i.e., the mathematica package for symbolic
semi-automatic evaluation of Feynman diagrams. In the present
report, the non-logarithmic loop corrections to the $q$-$\bar q$
annihilation cross section are taken into account by applying the
effective K-factor with a particular scale choice of ${\mu ^2} =
p_T^{4/3}{M^{2/3}}$ as it was done, for example, in the references
\cite{19,lipatov2011,field}, i.e.,
\[K = \exp [{C_F}\frac{{{\alpha _s}({\mu ^2})}}{{2\pi }}{\pi ^2}]\]
where ${p_T}$, ($p_T=\mid \vec{p_{1t}}+\vec{p_{2t}} \mid$, see
above the equation (\ref{9})), is the transverse momentum of
produced dilepton and ${C_F}$ is the color factor. To calculate the
multidimensional integration, the VEGAS routine \cite{vegas} is
used. The differential cross sections at several center of mass
energies, i.e., 1.96, 7 and 8 TeV as a function of the dilepton
invariant mass ($M$), rapidity ($y$), transverse momentum ($p_T$)
and the variable $\phi _\eta ^*$ \cite{phi1,phi2,phi3,phi4}, i.e.,
\[\phi _\eta ^* = \tan (\frac{{{\phi _{acop}}}}{2}){[cos(\frac{{\Delta \eta }}{2})]^{ - 1}},\]
are calculated, with ${\phi _{acop}} = \pi  - \left|
{\Delta \phi } \right|$, where ${\Delta \eta }$ and ${\Delta \phi }$ are the
pseudorapidity and azimuthal angles differences between the produced
leptons, as well, respectively. The variable $\phi _\eta ^*$ is correlated
to the quantity ${{\left| {{p_T}} \right|} \mathord{\left/
 {\vphantom {{\left| {{p_T}} \right|} M}} \right.
 \kern-\nulldelimiterspace} M}$ and both of them probes the same physics as the
dilepton transverse momentum, but it gives a better experimental
resolution \cite{56,57,58}.
\subsection{Results presentations}
\label{2} The results of above numerical calculations are compared
with the experimental data of  DY at the  Tevatron and LHC
laboratories with the total center of mass  energy $\sqrt s  = 1.8$
 TeV  and $\sqrt s =7$ and  $8$ TeV, respectively. We use the data
from different groups such as the CDF, CMS, ATLAS and LHCb
collaborations. The available pQCD predictions are also presented in
each figure.

The above comparisons are demonstrated in the figures 1 to 10 as
follows:
\\
(1): In all of the figures, the numerical results related to the KMR
UPDF are shown in the left panels in which  the dash, dotted-dash
and dotted histograms correspond to the contribution of individual
sub-processes, i.e., $q^* + \bar q^* \to {{{\gamma ^*}} \mathord{\left/
 {\vphantom {{{\gamma ^*}} {Z \to {l^ + } + {l^ - }}}} \right.
 \kern-\nulldelimiterspace} {Z \to {l^ + } + {l^ - }}}$, $q^* + {g^*} \to {{{\gamma ^*}} \mathord{\left/
 {\vphantom {{{\gamma ^*}} {Z \to {l^ + } + {l^ - } + q}}} \right.
 \kern-\nulldelimiterspace} {Z \to {l^ + } + {l^ - } + q}}$ and $q^* + \bar q^* \to {{{\gamma ^*}} \mathord{\left/
 {\vphantom {{{\gamma ^*}} {Z + g \to {l^ + } + {l^ - }}}} \right.
 \kern-\nulldelimiterspace} {Z + g \to {l^ + } + {l^ - }}} + g$,
 respectively. The shaded bands indicate the
corresponding uncertainty (${1\over 2}\leq \zeta\leq 2$) due to the
hard scale variation with   KMR UPDF  in cross sections evaluation.
Unlike the present report,    most of the previous phenomenological
works with the semi-KMR UPDF for the DY differential cross sections
did not present the contribution of each sub-process in their final
results and also, did not take  into account the contribution of the
third sub-process, assuming the possible double counting
\cite{lipatov2011,reggeiz1} between the first and the third
sub-processes. However, according to   our previous reports
\cite{18,diphoton}, we do not believe that there is any
double-counting among the first and the third sub-processes. This
point will be discussed in section \ref{3}.
\\
(2): In the right panels of each figure, the results of  different UPDF schemes applications, namely KMR, LO-MRW and NLO-MRW, in the
differential cross sections are shown by the solid, dash and
dotted-dash histograms, respectively,  for the possible comparisons.
\\
(3): The figures 1, 2-4, 5-8 and 9-10, demonstrate the DY
differential cross sections versus the diplepton invariant mass
($M$), the transverse momentum ($p_T$), the variable $\phi _\eta ^*$
and the rapidity ($y$), respectively (see the caption of each figure
for more details).
\subsection{Discussions}
\label{3} First, we generally start by  analyzing  the calculated
cross sections related to the medium and high center of mass
energies for $\sqrt s  = 1.8$ and $7$  TeV.  Although the results
show that the KMR UPDF describe reasonably the wide range of data of
Tevatron and LHC, but for the two sets of differential cross section
data which are in terms of $p_T$ and $\phi _\eta ^*$ parameters,
this is not the case. Indeed, in these two cases, the input NLO-MRW
UPDF describe the data better than other schemes. To be sure about
this conclusion,  we try to include the newer data from the ATLAS
collaboration at $\sqrt s = 8$ $TeV$. The results of this double
check are presented in the different panels of  figures 2-8. The
final comparisons, as we will be discussed below, indicates that
among the three different schemes, i.e., KMR, LO-MRW and NLO-MRW
UPDF, on average the NLO-MRW one is more suitable for describing
the experimental data and it confirms other groups  report of NNLO
pQCD calculations \cite{60}. We should point out here that in our
previous works, e.g., the reference \cite{diphoton}, because of the
possible fragmentation effects, the KMR and LO-MRW  had a better
agreement to the data.

The results of      double and single   differential cross sections
${{d\sigma}\over {dMdy}}$ and ${{d\sigma}\over {dM}}$ versus   the
invariant mass of the dilepton  are compared to the experimental
data at $\sqrt s = 1.8$ and $7$ TeV are shown in the figure 1, the panels
(a) and (b) and (c) to (f), respectively. In  the panels (a) to (d) of
this figure, as it is expected, the Z boson mass peak is observed
around the $M=91$ GeV. In these panels, it is clear that in the
small M region $(M<M_z)$ which corresponds to the medium and large
$p_T$, the contribution of LO $q$-$\bar q$ sub-process to the cross
section is less than the other ones. However, by increasing the
invariant mass of dilepton, its effect become larger than the other
sub-processes. In the panels (b), (d) and (f), the comparison
between
 all three approaches, i.e., KMR, LO-MRW and NLO-MRW, are presented and a similar behavior spatially at the
  Z boson mass region is observed. In the panel (d), it is clear that the results of three schemes are more
  or less the same, but in the panel (f) the KMR one shows more agreement with the experimental
  data. Our results in the panels (c) to (f) are also close to
  those of PYTHIA \cite{pythia1} and SHERPA \cite{d/dmATLAS}, especially in the small $M$ regions,
 but the reggeized \cite{reggeiz1,reggeiz2} model is below our
 predictions.   According to these panels,
  although our results show a overestimate and underestimate in the low and high dilepton invariant
  mass region, the uncertainty bands of our calculations cover the experimental data. In addition, the results of the SOC and the KaTie parton-level event generator are presented in panel (f) of this figure as well. It is clear that there is not any significant difference between these results.

In the figures 2-4, the normalized differential cross-sections of DY
   as a function of $p_T$, at $\sqrt s  = 7$ and $8$
TeV are compared to the CMS and ATLAS collaborations data. As it is
expected,   in all of the three figures, the first subprocess has
the main contribution while as we go to the higher center of mass
energy, i.e., $8$ TeV, the second subprocess also becomes important,
especially  with the increase of   dilepton mass (see the panels
(e) and (g) of the figure 2). In this figure, the results of   applying the SOC and the KaTie parton-level event generator are compared in the panels (d) and   (h), respectively. It is observed that
in   most of the regions, there isnot significant difference between the two schemes. Also in the figure 3 and 4 in which the
rapidity is increased, only the second subprocess   is sensitive to
the rapidity only in small $p_T$. However in the large $p_T$ region
($p_T>10$) the contribution of second subprocess, i.e., $q$-$g$,
becomes enhanced and in the middle of $p_T$ region only the first
and the second    sub-processes are in the same order. Now by
considering the above three UPDF schemes, one can find that for
$p_T<40 $, they behave very similar, while they are separated from
each other such that the LO-MRW and NLO-MRW
 are the upper and lower band of KMR, respectively. The experimental data also pass through the AOC band, see
the  reference \cite{diphoton}. A comparison between our results and
the parton level Monte Carlo programs
  such as PYTHIA and SHERPA are also presented. According to these panels,
  although our results show a overestimate and underestimate in the low and high dilepton invariant
  mass region, the uncertainty band of our calculations covers the experimental data.

In the figures 5-8, the normalized differential cross-sections of DY
at LHC as a function of the variable $\phi _\eta ^*$ and different
experimental conditions on the dilepton rapidity and invariant mass
are presented. According to these figures (beside the figure 5), it is
clear that in the small $\phi _\eta ^*$ region, which corresponds to
the back-to-back
 leptons, the contribution of the first and the second  sub-processes are dominated and approximately in the same
 order. But in the figure 5, which is demonstrated for the small mass
 interval, all of the three sub-processes are in the same order for the small $\phi _\eta ^*$
 region. On the other hand for $\phi _\eta ^*> 0.01$ the three
 sub-processes are separated and as it is expected   contribution
 of the  first sub-process becomes enhanced and ${{d\sigma_1}\over {\sigma d\phi _\eta ^*}}>
 {{d\sigma_2}\over {\sigma d\phi _\eta ^*}}\gg{{d\sigma_3}\over {\sigma d\phi _\eta
 ^*}}$.
 In the right panels of these figures, the same conclusion as above
 can be made about the effect of different UPDF schemes in which up
 to $\phi _\eta ^*< 0.1$ they behave the same, and for larger  $\phi _\eta
 $ as we discussed before, the NLO-MRW UPDF cross section
 calculations predict closer results to the corresponding data. The
 AOC and uncertainty bands   approximately cover the ATLAS
 collaboration data.

The differential cross sections of DY with respect to the rapidity of
dilepton versus y are plotted in the various panels of the figures 9
and 10. It is observed that on average the second sub-process ($qg$)
is dominant, especially in the mid-y region, compared to the
other two sub-processes. On the other hand by comparing the right
panels of these figures, one can conclude that again the NLO-MRW and
KMR schemes give closer results to experimental data with respect to
LO-MRW procedure.

In addition, our results are slightly different from the reference
\cite{reggeiz1} as we use different PDF, UPDF and method. Indeed, we
use the original method introduced by Kimber et al and consider the
correct form of the normalization equation (\ref{NOR}).

It is notable that the redefined form of normalization equation as
  $xa(x,\mu^2) \simeq \int^{\mu^2} {{dk_t^2}} f(x,k^2_t,\mu^2)$ without
  the factor $1/{k_t^2}$ does not lead to the collinear form of cross section
  after integrating over $k_t^2$. In the reference \cite{reggeiz1} the unpolarized
DY in the pp collisions is investigated at the LHC energies by CCFM
and semi-KMR within the reggeized quark formalism
\cite{reggeiz1,reggeiz2} to be sure about the gauge invariance of
matrix elements. However, as we discussed in our previous work
\cite{diphoton}, the  gauge invariance is guaranteed because of
applying the small-x-approximation in our calculations. As we
pointed out before, in the figures 1 (panel f) and  2 (panel b),   our results are
compared with those of references \cite{reggeiz1,reggeiz2}. On the
other hand, our results are compared with those of   PYTHIA
\cite{CDFM} (figure 1 (panels c-f)), SHERPA \cite{d/dmATLAS} (figures
1 ( panels e-f) and 3 ( panels a,c,e)),
 FEWZ \cite{LHCb7000} (figure 7 (panels g-h)) and RESBOS \cite{ResBos2} (figures 4 (panel e) and 7 (panel e)). It is observed that
in the regions in which the higher order calculations are not
important our results are similar to those of  pQCD. However, in
some parts, spatially for high $p_T$ and $\phi _\eta ^*$, the
  NLO-MRW UPDF scheme shows slightly different  behavior with respect to the data and the
pQCD methods.

In several papers such as the reference  \cite{60}, the authors
denote  that the
 description of two observables, including the $p_T$ and $\phi _\eta ^*$ distributions,
 are
  improved, if the higher order perturbative contributions are taken into account, which is in agreement with the cross check we  performed.
On the other hand as the scale of energy increases, for the LHC
energies, the $q$-$g$
  sub-process has the largest contribution to the differential cross section in the most intervals
  of $p_T$ and  $\phi _\eta ^*$, as it is expected.

We also checked the interference effect between the $\gamma^*$ and
$Z$ in the cross sections and find out that the interference is ignorable in
all regions.

Finally, we would like to point out that there is a new CMS measurement on the differential cross sections of the Z boson production in the P-P collisions \cite{CMS2019}. In the figures 7 (in terms of $p_T^Z$) and 8 (in terms of $\phi^*_\eta$ of dilepton) of this report, the CMS data are compared to the theoretical works presented in the reference \cite{R1,R2,R3,R4}, in which the UPDF (the so called transverse momentum  dependent distribution  functions (TMD)) are calculated, using the Parton Branching (PB) model. In this PB TMD model, the resummation to NLL accuracy, the fixed-order results at NLO, and the nonperturbative contributions are taken into account. The PB TMD results can predict the data well at low  $p_T^Z$, but deviates from the measurements at
high  $p_T^Z$, because of missing contributions from Z+jets matrix element calculations.  Furthermore, in the present work, we do not use the LHAPDF \cite{LHAPDF} or TMDlib \cite{TMDlib}  repositories, but we hope in our future reports, we can analyze the difference between the applications of  present PDF and UPDF with   those can be generated through  LHAPDF and TMDlib repositories.
\section{Conclusions}
We investigated the lepton pair production in the $p$-$p$ and
$p$-$\bar p$ collisions within the framework of
${k_t^{}}$-factorization approach. We used the transverse momentum
dependent parton distribution functions of three different
prescriptions, i.e., KMR, LO-MRW and NLO-MRW. We calculated the
matrix element square for the three different sub-processes among
which the matrix element square for the $q$-$\bar q$ in the  NLO
level is rarely taken into account. We calculated several
differential cross sections in terms of    the dilepton invariant
mass, transverse momentum and rapidity, as well as the angular
correlation between produced leptons of the Drell-Yan process. In
addition, we obtained the uncertainty band for the cross section
distribution in the case of KMR by changing the scale factor as
described in the section III. We considered the contribution of each
sub-processes separately based on the off-shell and massless quarks.
We found that although some of the  results show that using the KMR
framework, rather than LO-MRW and NLO-MRW schemes, represents more
agreement with the experimental data, in the case of $p_T$ and $\phi
_\eta ^*$ probing  the NLO-MRW gave  better predictions. It is shown that the AOC and SOC constraints give similar results and our direct calculations of the off shell matrix elements and the method of integration for evaluation of the cross section give   the same prediction as those of KaTie parton-level event generator.  

Finally, in this work we consider   the
renormalization and factorization scales to be equal, i.e., $\mu _R =
\mu _F = \zeta M$, in which,
$M$ is  the
invariant mass of produced dilepton  and $\zeta$  can  
vary  from 1/2 to 2, to estimate the scale uncertainties of our
calculations. However as stated in the reference \cite{CMS2019}, one can vary each scale independently. Beside this it is possible to find the uncertainty, which come through the implementation of PDF through UPDF. But it should not be as large as the uncertainty effect due to the  variation of renormalization and factorization scales. We hope to verify these effects in our future reports.
\begin{acknowledgements}
M. Modarres and R. Taghavi would like to acknowledge the research
support of University of Tehran and   the Iran National Science
Foundation (INSF) for their  grants.
\end{acknowledgements}
\appendix
\section{ }
\label{a}  The matrix elements of three sub-processes can be
presented as follows:

\begin{equation}
{\cal M}_1^\gamma  = i{e^2}{e_q}{{\bar \upsilon
}_{s1}}({k_2}){\gamma ^\mu }{u_{s2}}({k_1})\frac{{{g_{\mu \nu
}}}}{s}{{\bar u}_{r1}}({p_1}){\gamma ^\nu }{\upsilon _{r2}}({p_2}),
\end{equation}

\[M_1^Z = i\frac{{g_w^2}}{{4{{\cos }^2}{\theta _w}}}{{\bar \upsilon }_{s1}}({k_2}){\gamma ^\mu }(C_V^q - C_A^q{\gamma ^5}){u_{s2}}({k_1}) \times \]
 \begin{equation}
 \times ({g_{\mu \nu }} - \frac{{{{({k_1} + {k_2})}_\mu }{{({k_1} + {k_2})}_\nu }}}{{m_Z^2}})\frac{{{{\bar u}_{r1}}({p_1}){\gamma ^\nu }(C_V^e - C_A^e{\gamma ^5}){\upsilon _{r2}}({p_2})}}{{(s - m_Z^2 - i{m_Z}{\Gamma _Z})}},
 \end{equation}

\[{\cal M}_2^\gamma  =  - {e^2}{e_q}{g_s}{t^a}{\varepsilon _\mu }({k_2}){{\ u}_{s1}}({k_1})({\gamma ^\mu }\frac{{{{\hat k}_1} + {{\hat k}_2}}}{s}{\gamma ^\nu } + {\gamma ^\nu }\frac{{ - {{\hat k}_2} + {{\hat p}_3}}}{{{{( - {k_2} + {p_3})}^2}}}
{\gamma ^\mu }){ {\bar u}_{s2}}({p_3})\times\]
  \begin{equation}
 \times \frac{{{g_{\mu \nu }}}}{{{{({p_1} + {p_2})}^2}}}{{\bar u}_{r1}}({p_1}){\gamma ^\rho }{\upsilon _{r2}}({p_2}),
   \end{equation}
  \[{\cal M}_2^Z =  - \frac{{g_w^2{g_s}}}{{4{{\cos }^2}{\theta _w}}}{t^a}{\varepsilon _\mu }({k_2})\]
\[{{\ u}_{s1}}({k_1})({\gamma ^\mu }(C_V^q - C_A^q{\gamma ^5})\frac{{{{\hat k}_1} + {{\hat k}_2}}}{s}{\gamma ^\nu }
+ {\gamma ^\nu }\frac{{ - {{\hat k}_2} + {{\hat p}_3}}}{{{{( - {k_2}
+ {p_3})}^2}}}{\gamma ^\mu }(C_V^q - C_A^q{\gamma ^5})){{\bar
u}_{s2}}({p_3})\times\]
 \begin{equation}
 \times ({g_{\rho \nu }} - \frac{{{{({p_1} + {p_2})}_\rho }{{({p_1}
 + {p_2})}_\nu }}}{{m_Z^2}})\frac{{{{\bar u}_{r1}}({p_1}){\gamma ^\rho }(C_V^q - C_A^q{\gamma ^5}){\upsilon _{r2}}({p_2})}}{{{{({p_1} + {p_2})}^2}
 - m_Z^2 - i{m_Z}{\Gamma _Z}}},
\label{7p}
 \end{equation}

\[{\cal M}_3^\gamma  =  - {e^2}{e_q}{g_s}{t^a}{\varepsilon _\mu }({p_3}){u_{s1}}({k_1})({\gamma ^\mu }\frac{{{{\hat k}_1}
 - {{\hat p}_3}}}{{{{({{\hat k}_1} - {{\hat p}_3})}^2}}}{\gamma ^\nu } +
 {\gamma ^\nu }\frac{{ - {{\hat k}_2} + {{\hat p}_3}}}{{{{( - {k_2} + {p_3})}^2}}}{\gamma ^\mu }){{\bar \upsilon }_{s2}}({k_2}) \times \]

 \begin{equation}
\times \frac{{{g_{\nu \rho }}}}{{{{({p_1} + {p_2})}^2}}}{{\bar
u}_{r1}}({p_1}){\gamma ^\rho }{\upsilon _{r2}}({p_2}),\label{eq7pp}
\end{equation}

\[{\cal M}_3^Z =  - \frac{{g_w^2{g_s}}}{{4{{\cos }^2}{\theta _w}}}{t^a}{\varepsilon _\mu }({p_3}){u_{s1}}({k_1}) \times \]

\[ \times ({\gamma ^\mu }(C_V^q - C_A^q{\gamma ^5})\frac{{{{\hat k}_1}
- {{\hat p}_3}}}{{{{({{\hat k}_1} - {{\hat p}_3})}^2}}}{\gamma ^\nu
} + {\gamma ^\nu }\frac{{ - {{\hat k}_2} + {{\hat p}_3}}}{{{{( -
{k_2} + {p_3})}^2}}}{\gamma ^\mu }(C_V^q - C_A^q{\gamma ^5})){{\bar
\upsilon }_{s2}}({k_2}) \times \]
 \begin{equation}
 \times ({g_{\rho \nu }} - \frac{{{{({p_1} + {p_2})}_\rho }{{({p_1}
 + {p_2})}_\nu }}}{{m_Z^2}})\frac{{{{\bar u}_{r1}}({p_1}){\gamma ^\rho }(C_V^q - C_A^q{\gamma ^5}){\upsilon _{r2}}({p_2})}}{{{{({p_1}
 + {p_2})}^2} - m_Z^2 - i{m_Z}{\Gamma _Z}}}, \end{equation}
where $s = {({k_1} + {k_2})^2}$  and the electron and quark (fractional)
electric charges are denoted by $e$ and ${e_q}$. Other notations are
the same as the reference \cite{lipatov2011}.
\section{  }
\label{b} The NLO splitting functions are defined as [21]:

    \begin{equation}
\tilde{P}_{ab}^{(0+1)}(z) = \tilde{P}_{ab}^{(0)}(z) + {\alpha_S
\over 2\pi}
    \tilde{P}_{ab}^{(1)}(z),
    \label{eq12p}
    \end{equation}
with
    \begin{equation}
\tilde{P}_{ab}^{(i)}(z) = P_{ab}^{(i)}(z) - \Theta (z-(1-\Delta))
\delta_{ab} F^{(i)}_{ab} P_{ab}(z),
    \label{eq13}
    \end{equation}
where $i= 0$ and $1$ stand for the $LO$ and the $NLO$, respectively.
$\Delta$ can be defined as \cite{MRW}:
    $$ \Delta = {k\sqrt{1-z} \over k\sqrt{1-z} + \mu}.$$ and we have:
\begin{equation}
F_{qq}^{(0)} = {C_F},
 \label{eq12pp}
    \end{equation}
\begin{equation}
F_{qq}^{(1)} =  - {C_F}({T_R}{N_F}\frac{{10}}{9} + {C_A}(\frac{{{\pi
^2}}}{6} - \frac{{67}}{{18}})),
 \label{eq12ppp}
    \end{equation}

\begin{equation}
F_{gg}^{(0)} = 2{C_A},
 \label{eq12pppp}
    \end{equation}

\begin{equation}
F_{gg}^{(1)} =  - 2{C_F}({T_R}{N_F}\frac{{10}}{9} +
{C_A}(\frac{{{\pi ^2}}}{6} - \frac{{67}}{{18}})),
 \label{eq12ppppp}
    \end{equation}
\begin{equation}
{P_{qq}}(z) = \frac{{(1 - {z^2})}}{{1 - z}},
 \label{eq12pppppppp}
    \end{equation}
\begin{equation}
{P_{gg}}(z) = \frac{z}{{(1 - z)}} + \frac{{(1 - z)}}{z} + z(1 - z),
 \label{eq127}
    \end{equation}

\newpage
  \includegraphics[width=\textwidth, height=15cm]{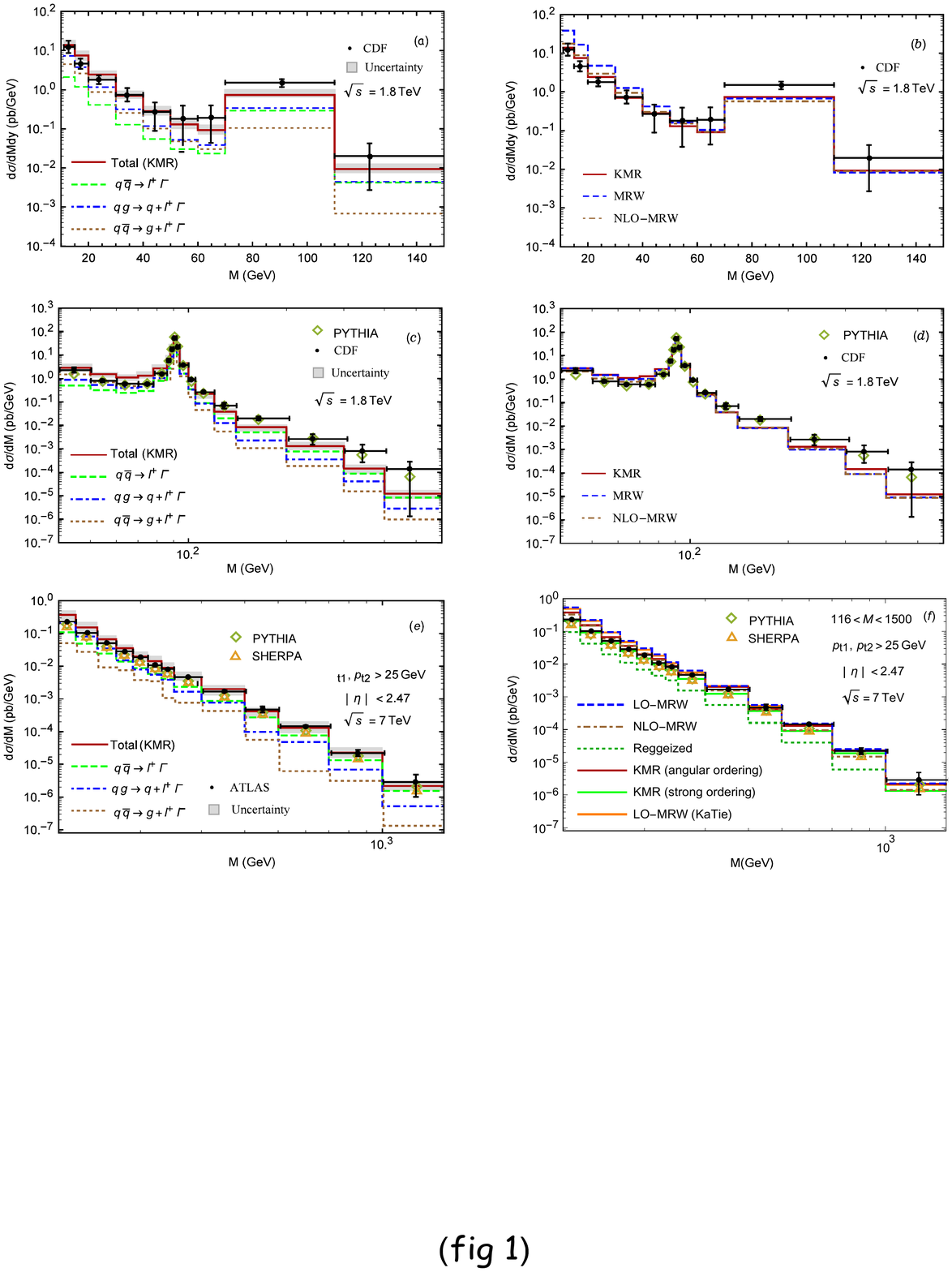}
Fig 1: The (double) differential cross-sections of DY at Tevatron and LHC as
a function of the dilepton invariant mass at $E_{CM} =1.8$ and $7$ TeV
compared to the CDF and ATLAS data \cite{CDFM,CDFMY,d/dmATLAS}. The numerical results
related to the $KMR$ UPDF are shown in left panels. The
contribution of $ q^* + \bar q^* \to {{{\gamma ^*}} \mathord{\left/
 {\vphantom {{{\gamma ^*}} {Z \to {l^ + } + {l^ - }}}} \right.
 \kern-\nulldelimiterspace} {Z \to {l^ + } + {l^ - }}
}$, $ q^* + {g^*} \to {{{\gamma ^*}} \mathord{\left/
 {\vphantom {{{\gamma ^*}} {Z \to {l^ + } + {l^ - } + q}}} \right.
 \kern-\nulldelimiterspace} {Z \to {l^ + } + {l^ - } + q}}
$ and
 $
q^* + \bar q^* \to {{{\gamma ^*}} \mathord{\left/
 {\vphantom {{{\gamma ^*}} {Z + g \to {l^ + } + {l^ - }}}} \right.
 \kern-\nulldelimiterspace} {Z + g \to {l^ + } + {l^ - }}} + g
$ sub-processes are presented by dash, dotted-dash and dotted
histograms. In the right panels, the results corresponding to the
KMR, LO and NLO-MRW UPDF are shown by solid, dashed and dotdashed
histograms respectively and compared with each other . The shaded
bands indicate the corresponding uncertainty for KMR calculations 
(see the text for details about the SOC and KaTie   results).\label{fig1}\\
\newpage 
  \includegraphics[width=\textwidth, height=19cm]{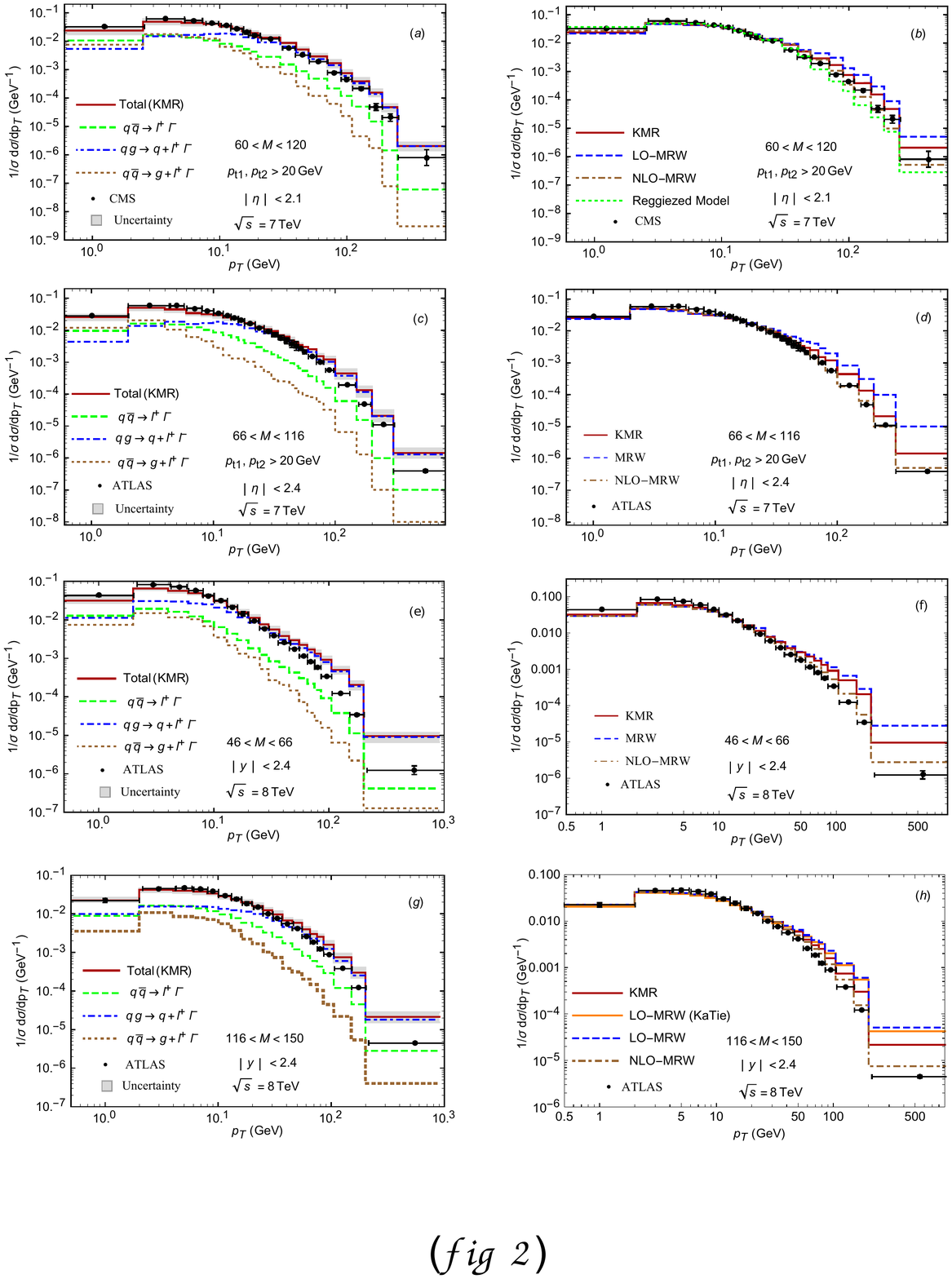}
Fig 2: The normalized differential cross-section of Drell-Yan lepton
pair production at LHC as a function of the dilepton transverse
momentum ,  at $E_{CM} =7$ and $8$ TeV compared to the
different collaborations data from the CMS \cite{20} and ATLAS
\cite{24, fig2panel-c}. The notation of all histograms is the same as in
the figure 1 (see the text for details about the SOC and KaTie results).\label{fig2}\\
   \includegraphics[width=\textwidth, height=19cm]{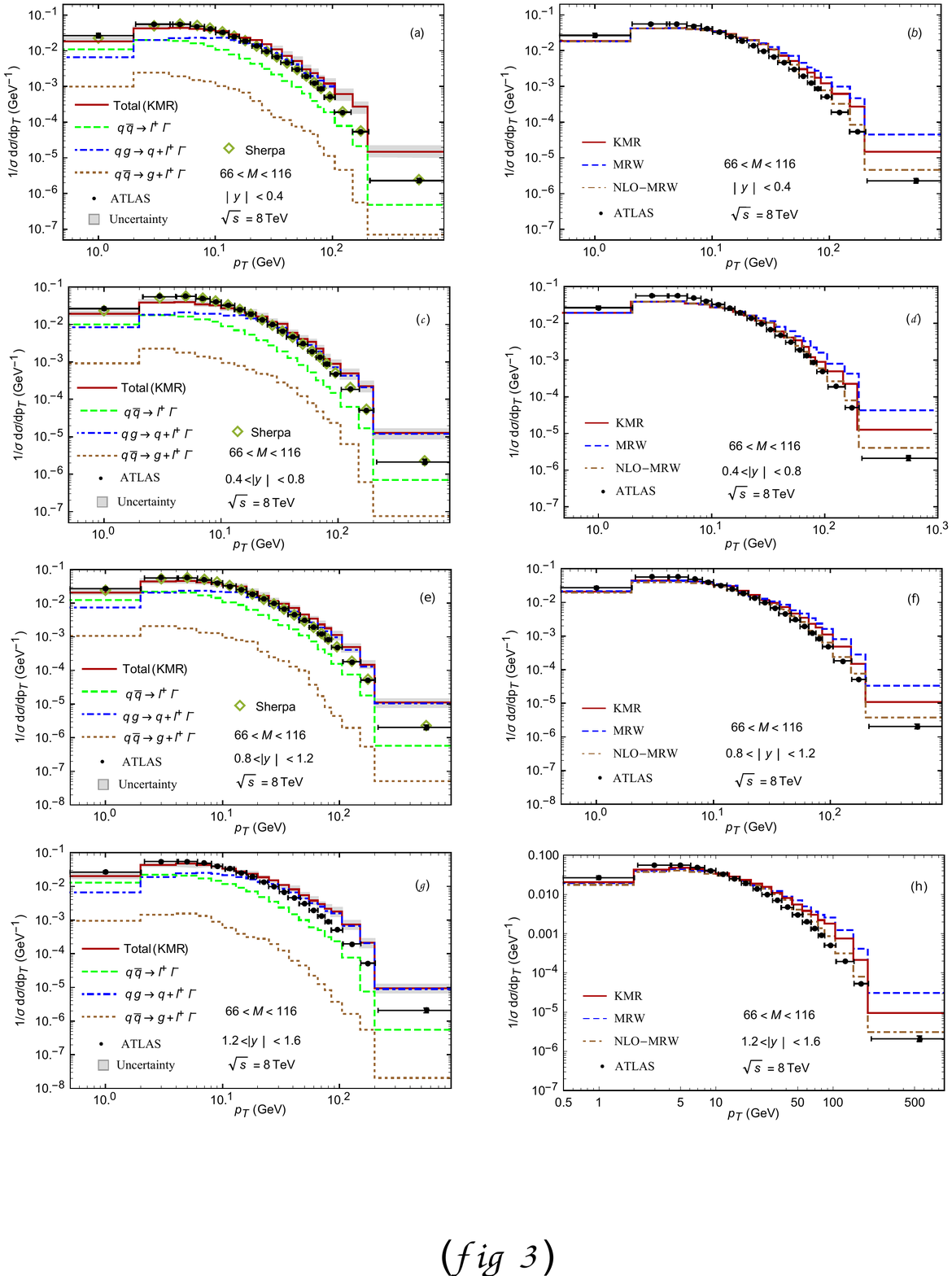}
Fig 3: The normalized differential cross-section of Drell-Yan lepton
pair production at LHC as a function of   dilepton
transverse momentum  at $E_{CM} =8$   TeV compared to the   ATLAS
data \cite{26}. The notation of all histograms is the same as
in the figure 1.\label{fig3}\\
  \includegraphics[width=\textwidth, height=19cm]{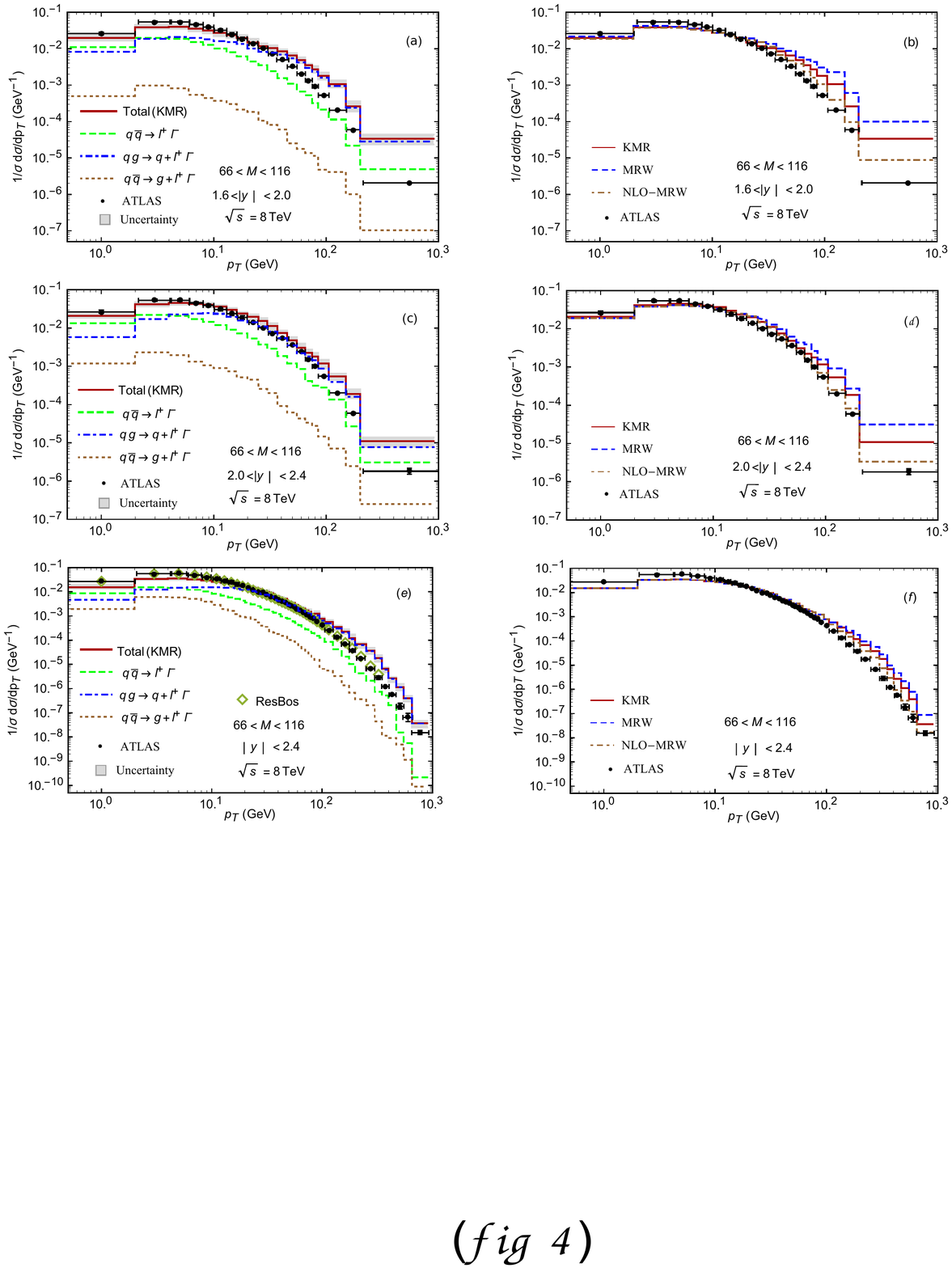} 
Fig 4: The differential cross-section of Drell-Yan lepton pair
production at LHC as a function of the dilepton transverse momentum  at $E_{CM}
=8$TeV compared to the ATLAS data \cite{26}.The notation of all
histograms is the same as in the figure 1. \label{fig4}\\
  \includegraphics[width=\textwidth, height=19cm]{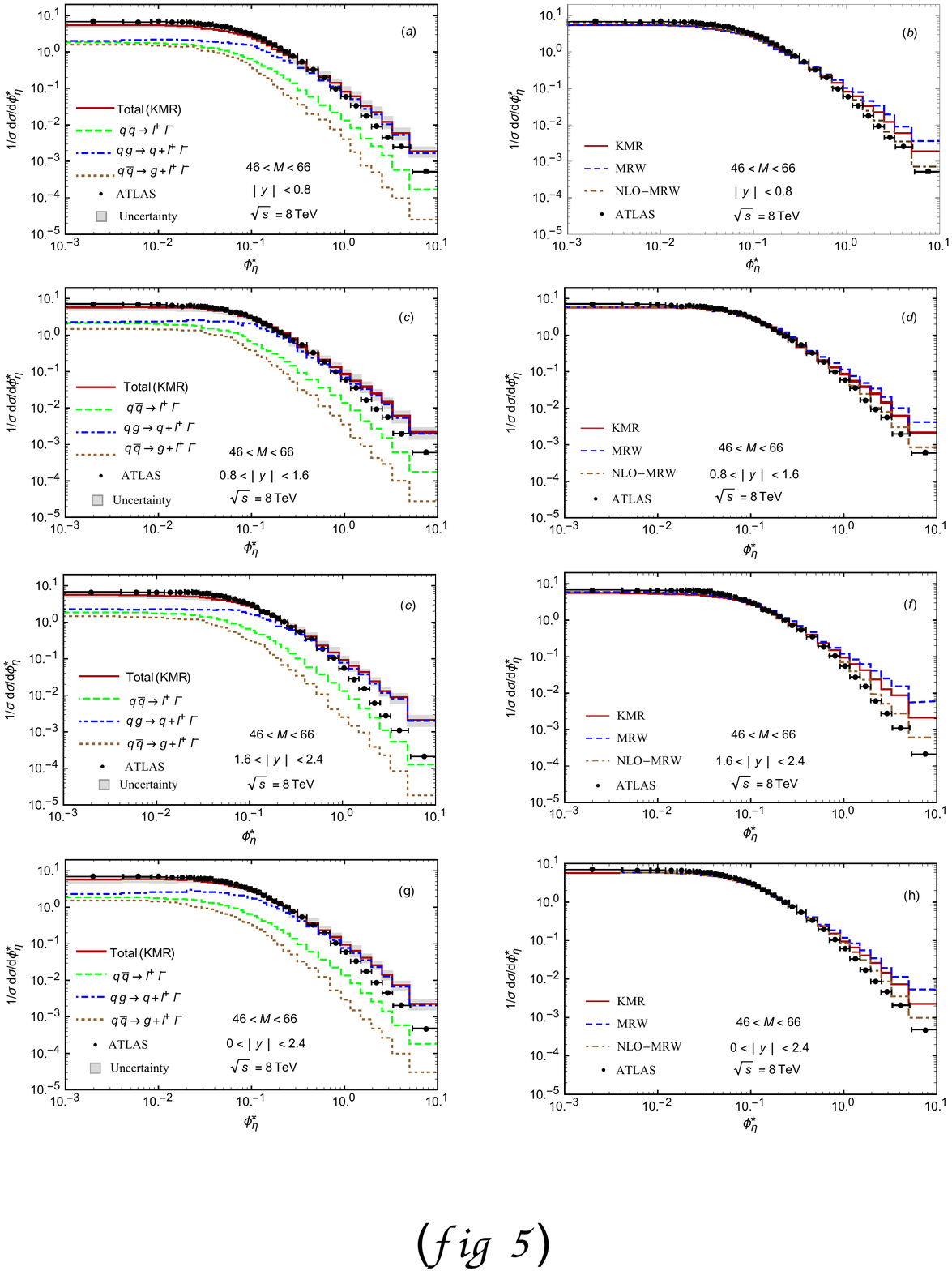}  
Fig 5: The normalized differential cross-section of Drell-Yan lepton
pair production at LHC as a function of $\phi _\eta ^*$ at $E_{CM}
=8$ TeV compared to the ATLAS data \cite{26}. The notation of
all histograms is the same as in the figure 1. \label{fig5}\\
  \includegraphics[width=\textwidth, height=19cm]{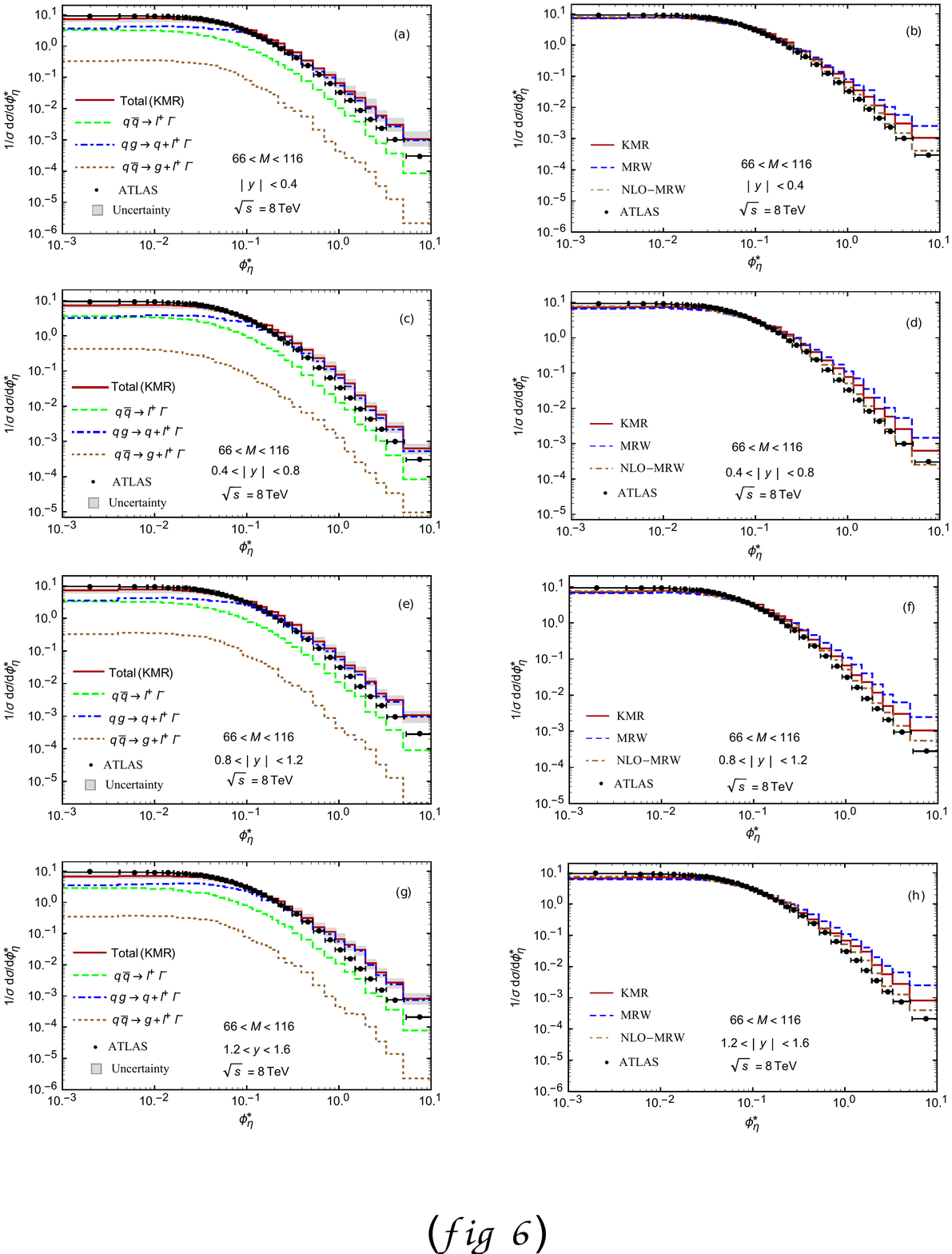} 
Fig 6: The normalized differential cross-section of Drell-Yan lepton
pair production at LHC as a function of $\phi _\eta ^*$ at $E_{CM}
=8$ TeV compared to the ATLAS data \cite{26}.  The otation of
all histograms is the same as in the figure 1. \label{fig6}\\
  \includegraphics[width=\textwidth, height=19cm]{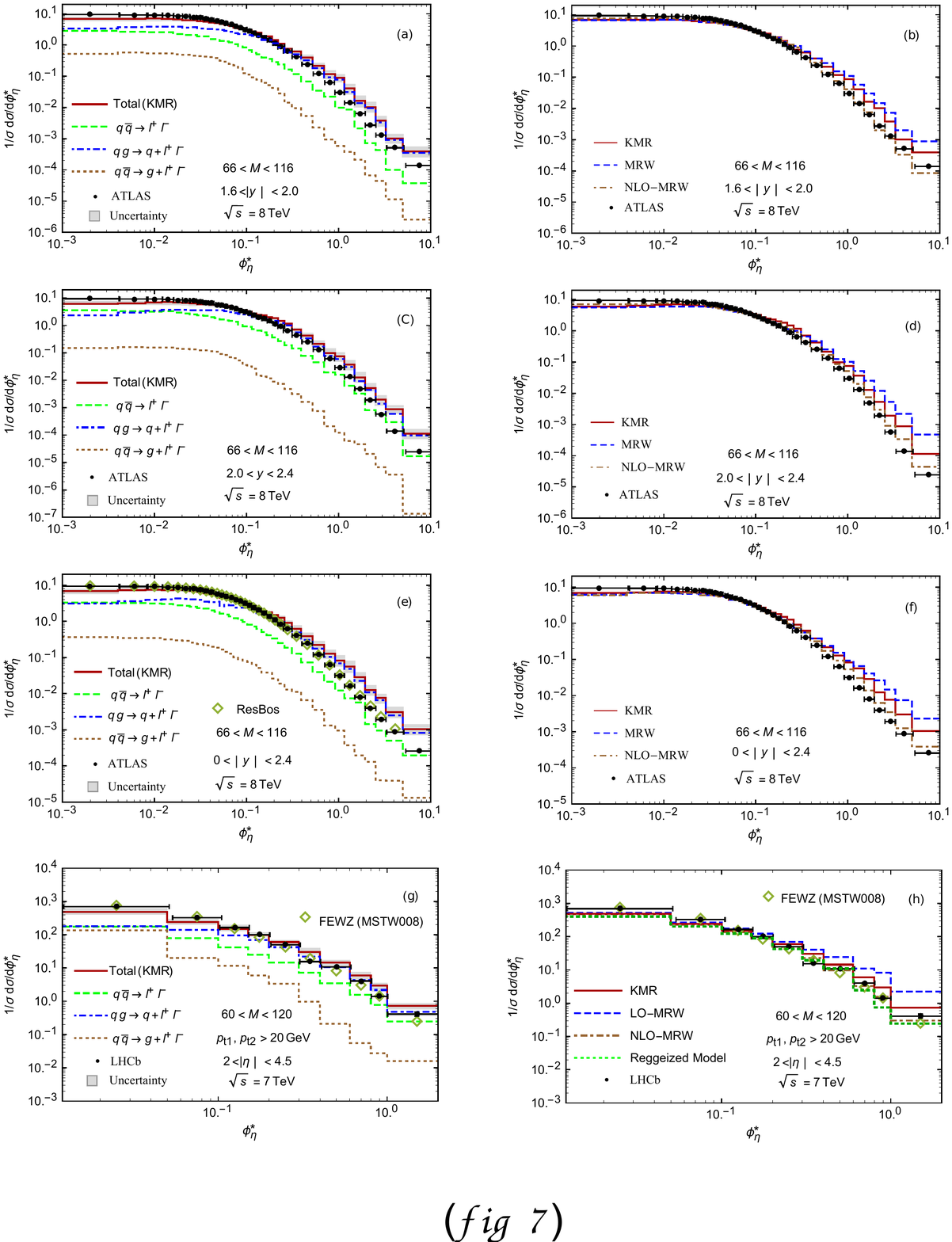} 
Fig 7: The normalized differential cross-section of Drell-Yan lepton
pair production at LHC as a function of $\phi _\eta ^*$ at $E_{CM}
=8$ TeV compared to the ATLAS data \cite{26}. The notation of
all histograms is the same as in the figure 1. \label{fig7}\\
  \includegraphics[width=\textwidth, height=19cm]{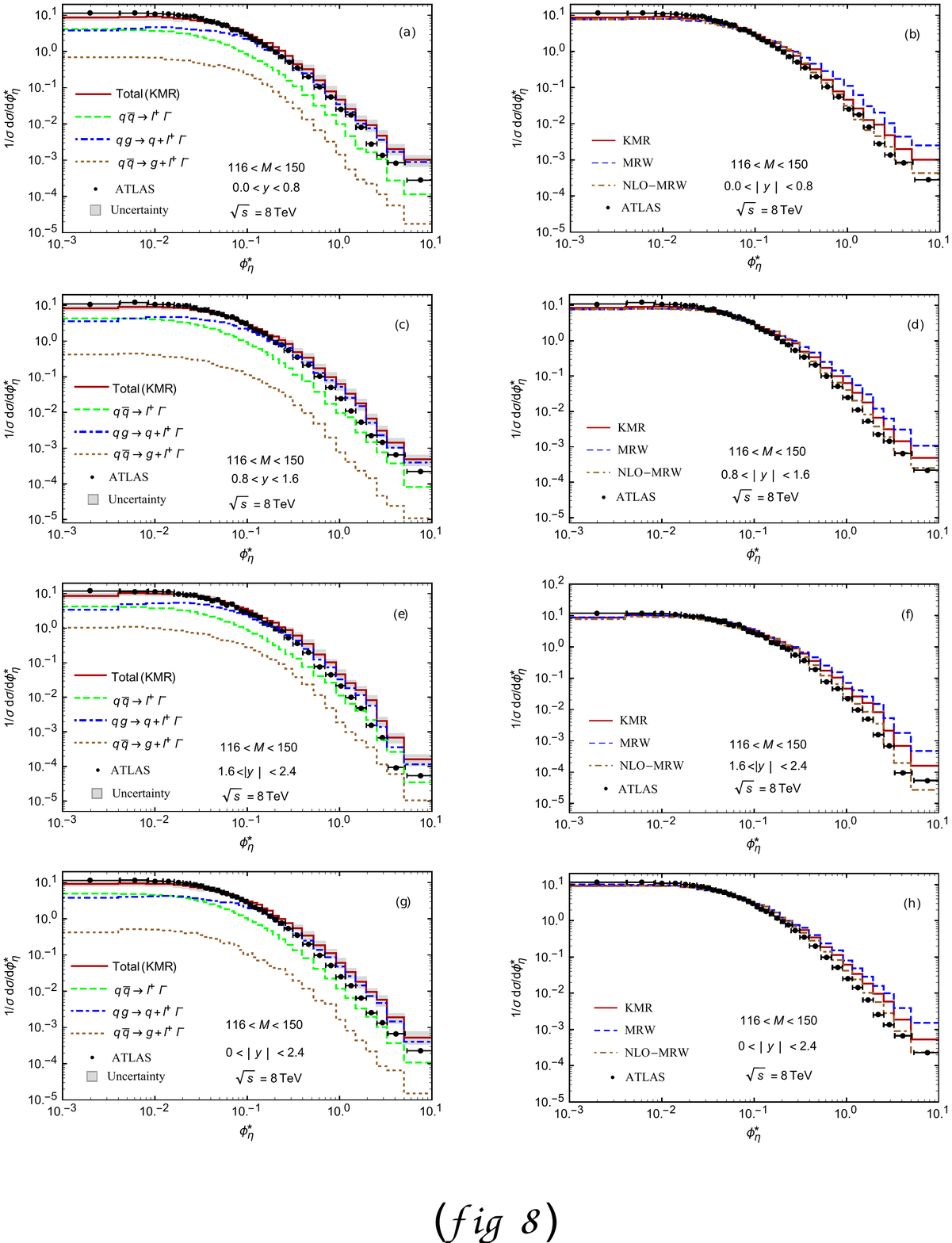} 
Fig 8: The normalized differential cross-section of Drell-Yan lepton
pair production at LHC as a function of $\phi _\eta ^*$ and dilepton
transverse momentums at $E_{CM} =8$ TeV compared to the ATLAS data
\cite{26}. The notation of all histograms is the same as in
the figure 1. \label{fig8}\\
   \includegraphics[width=\textwidth, height=19cm]{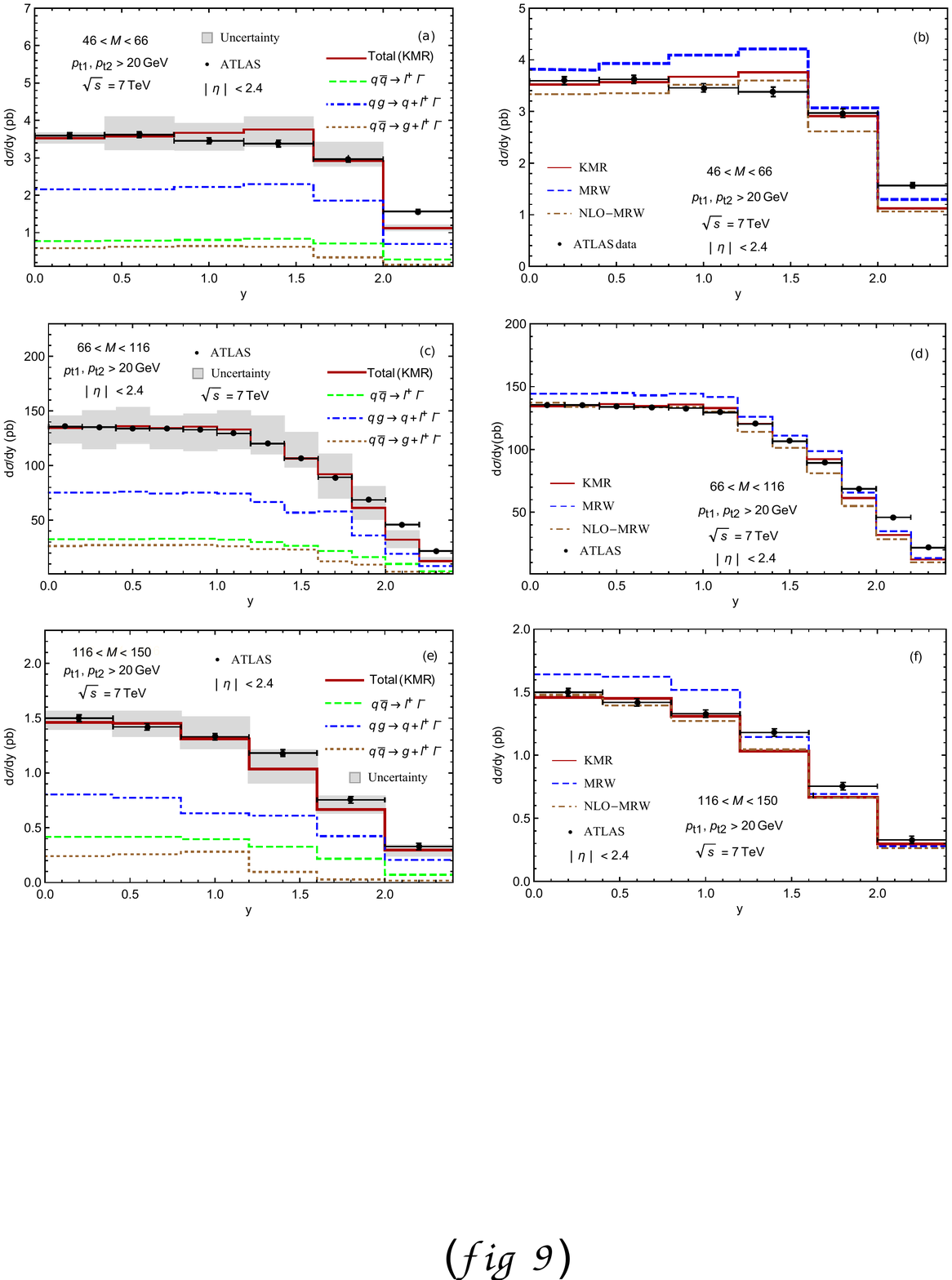}  
Fig 9: The normalized differential cross-section of Drell-Yan lepton
pair production at LHC as a function of the dilepton rapidity at $E_{CM}$ = $7$ TeV compared to the ATLAS data \cite{65}.
The notation of all histograms is the same as in
the figure 1. \label{fig9}\\
 \includegraphics[width=\textwidth, height=19cm]{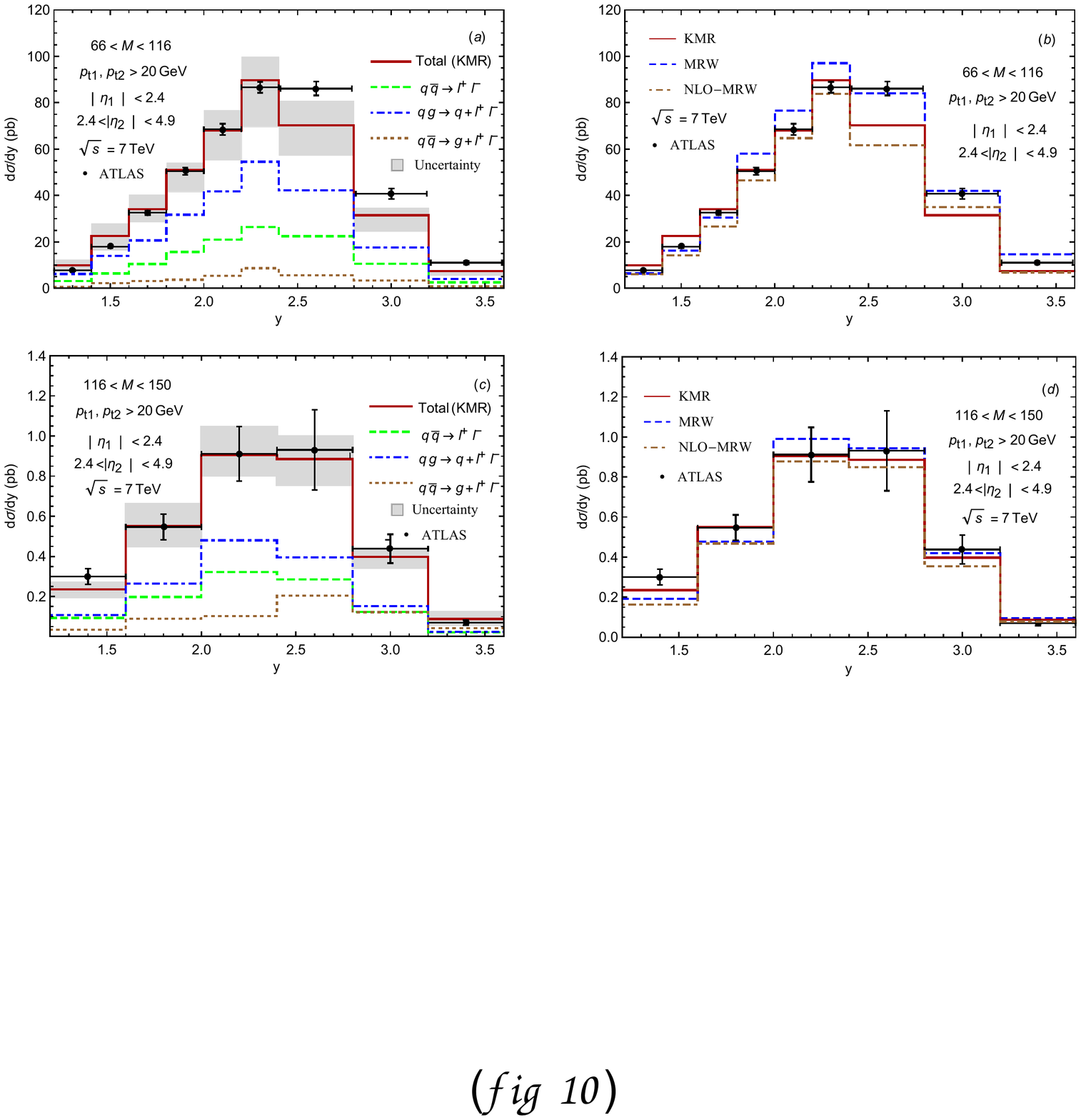} 
Fig 10: The normalized differential cross-section of Drell-Yan
lepton pair production at LHC as a function of the dilepton
rapidity at $E_{CM} =7$ TeV compared to the ATLAS data
\cite{65}. The notation of all histograms is the same as in the figure 1 .
\label{fig10}

\end{document}